\documentclass[twocolumn,traditabstract,longauth]{aa}
\usepackage{graphicx}

\begin{document}

\title{
Observation of H$_2$O in a strongly lensed {\it Herschel}-ATLAS \thanks{{\it Herschel} is an ESA space observatory with science instruments provided by European-led Principal Investigator consortia and with important participation from NASA.} 
source at z=2.3
\thanks{Colour figure only available in the electronic
form via http://www.edpsciences.org}}

\author{A. Omont\inst{1,2}
\and R. Neri\inst{3}
\and P. Cox\inst{3}
\and R. Lupu\inst{4}
\and M. Gu\'elin\inst{3}
\and P. van der Werf\inst{5}
\and A. Wei\ss\inst{22}
\and R. Ivison\inst{16}
\and M. Negrello\inst{6}
\and L. Leeuw\inst{19}
\and M. Lehnert\inst{18}
\and I. Smail\inst{11}
\and A. Beelen\inst{7}
\and J.E. Aguirre\inst{4}
\and M. Baes\inst{31}
\and F. Bertoldi\inst{8}
\and D.L. Clements\inst{9}
\and A. Cooray\inst{10}
\and K. Coppin\inst{35}
\and H. Dannerbauer\inst{12}
\and G. De Zotti\inst{13}
\and S. Dye\inst{14}
\and N. Fiolet\inst{1,2,7}
\and D. Frayer\inst{23}
\and R. Gavazzi\inst{1,2}
\and D. Hughes\inst{15}
\and M. Jarvis\inst{17}
\and M. Krips\inst{3}
\and M. Micha{\l}owski\inst{25}
\and E. Murphy\inst{28}
\and D. Riechers\inst{28}
\and A.M. Swinbank\inst{11}
\and P. Temi\inst{20}
\and M. Vaccari\inst{24}
\and A. Verma\inst{21}
\and J.D. Vieira\inst{28}
\and R. Auld\inst{14} 
\and B. Buttiglione\inst{26}
\and A. Cava\inst{32}
\and A. Dariush\inst{27,14}
\and L. Dunne\inst{17} 
\and S.A. Eales\inst{14}
\and J. Fritz\inst{26}
\and H. Gomez\inst{14}
\and E. Ibar\inst{16}
\and S. Maddox\inst{17}
\and E. Pascale\inst{14}
\and M. Pohlen\inst{14}
\and E. Rigby\inst{17}
\and D. Smith\inst{17}
\and A.J. Baker\inst{34}
\and J. Bock \inst{28,29}
\and C.M. Bradford \inst{28,29}
\and J. Glenn \inst{30}
\and A.I. Harris\inst{33}
\and K.S. Scott \inst{4}
\and J. Zmuidzinas \inst{28,29}
}
\institute{UPMC Univ Paris 06, UMR7095, Institut d'Astrophysique de Paris, 
F-75014, Paris, France
\and  CNRS, UMR7095, Institut d'Astrophysique de Paris, F-75014, Paris, France 
\and  Institut de Radioastronomie Millim{\'e}trique (IRAM), 300 rue de la Piscine, 38406 Saint-Martin d'H{\`e}res, France
\and  Department of Physics and Astronomy, University of Pennsylvania, Philadephia, PA 19104, US
\and  Leiden Observatory, Leiden University, Post Office Box 9513, NL - 2300 RA Leiden, The Netherlands 
\and  Department of Physics and Astronomy, The Open University, Walton Hall, Milton Keynes, MK7 6AA, UK
\and  Univ Paris-Sud and CNRS, Institut d'Astrophysique Spatiale, UMR8617, Orsay, F-91405, France
\and  Argelander Institut fur Astronomie, Universit Bonn, Auf dem H$\rm\ddot{u}$gel 71, 53121 Bonn, Germany
\and  Astrophysics Group, Physics Department, Blackett Lab, Imperial College London, Prince Consort Road, London SW7
\and  Department of Physics and Astronomy, University of California, Irvine, CA 92697, USA
\and  Institute for Computational Cosmology, Durham University, South Road, Durham DH1 3LE
\and  Laboratoire Astrophysique, Instrumentation et Mod$\rm\acute{e}$lisation Paris Saclay, Commissariat $\rm\grave{a}$l'$\acute{E}$nergie Atomique (CEA)/Direction des Sciences de la Mati$\rm\grave{e}$re (DSM) - CNRS - Universit$\rm\acute{e}$ Paris Diderot, Institut de recherche sur les lois fondamentales de l'Univers (Irfu)/Service d'Astrophysique, CEA Saclay, Orme des Merisiers, F-91191 Gif-sur-Yvette Cedex, France
\and  INAF-Osservatorio Astronomico di Padova, Vicolo Osservatorio 5, I-35122 Padova, Italy; and SISSA, Via Bonomea 265, I-34136 Trieste, Italy
\and  School of Physics and Astronomy, Cardiff University, The Parade, Cardiff, CF24 3AA, UK
\and  Instituto Nacional de Astrof\'{i}sica, \'{O}ptica y Electr\'{o}nica, Apartado Postal 51 y 216, 72000 Puebla, Mexico
\and  UK Astronomy Technology Center, Royal Observatory Edinburgh, Edinburgh, EH9 3HJ, UK
\and  School of Physics and Astronomy, University of Nottingham, University Park, Nottingham NG7 2RD, UK
\and  GEPI, Observatoire de Paris, CNRS, Universit\'e Paris Diderot, 5 Place Jules Janssen, 92190 Meudon, France
\and  SETI Institute, 515 North Whisman Avenue Mountain View CA, 94043, USA
\and   Astrophysics Branch, NASA Ames Research Center, MS 245-6, Moffett Field, CA 94035, USA
\and   Oxford Astrophysics, Denys Wilkinson Building, University of Oxford, Keble Road, Oxford, OX1 3RH,  UK
\and  Max-Plank-Institut für Radioastronomie, Auf dem Hügel 69, 53121 Bonn, Germany
\and National Radio Astronomy Observatory Post Office Box 2, Green Bank, WV 24944, USA 
\and  INAF-IASF Milano, via E. Bassini 15, 20133, Italy
\and  Institute for Astronomy, University of Edinburgh, Royal Observatory, Blackford Hill, Edinburgh EH9 3HJ
\and  University of Padova, Department of Astronomy, Vicolo Osservatorio 3, 35122 Padova, Italy
\and  School of Astronomy, Institute for Research in Fundamental Sciences (IPM),PO Box 19395-5746, Tehran, Iran 
\and  California Institute of Technology, Pasadena, CA 91125, USA
\and  Jet Propulsion Laboratory, Pasadena, CA 91109, USA
\and  University of Colorado, CASA 389-UCB, Boulder, CO 80303, USA
\and  Sterrenkundig Observatorium, Universiteit Gent, Krijgslaan 281-S9, 9000 Gent, Belgium 
\and Departamento de Astrof\'{\i}sica, Facultad de CC. F\'{\i}sicas, Universidad Complutense de Madrid, E-28040 Madrid, Spain
\and  Department of Astronomy, University of Maryland, College Park, MD 20742, USA harris@astro.umd.edu
\and  Department of Physics and Astronomy, Rutgers, The State University of New Jersey, Piscataway, NJ 08854-8019, USA
\and  Department of Physics, McGill University, Ernest Rutherford Building, 3600 Rue University, Montr\'{e}al, Qu\'{e}bec, H3A 2T8, Canada
}



\abstract{The {\it Herschel} survey, H-ATLAS, with its large areal coverage, has recently discovered a number 
of bright, strongly lensed high-$z$ submillimeter galaxies. The strong magnification makes it possible to study molecular
species other than CO, which are otherwise difficult to observe in high-$z$ galaxies. Among the lensed galaxies 
already identified by H-ATLAS, the source J090302.9-014127B (SDP.17b) at $z=2.305$ is remarkable due to its excitation 
conditions and a tentative detection of the H$_2$O 2$_{02}$-1$_{11}$ emission line (Lupu et al.\ 2010). 
We report observations of this line in SDP.17b using the IRAM interferometer equipped with its new 277--371\,GHz 
receivers. 
The H$_2$O line is detected  at a redshift of $z=2.3049\pm0.0006$, with a flux of 7.8$\pm$0.5\,Jy km s$^{-1}$ 
and a FWHM of $\rm 250 \pm 60 \,km \, s^{-1}$. The new flux is 2.4 times weaker than the previous
tentative detection, although both remain marginally consistent within 1.6$\sigma$. The intrinsic line 
luminosity and ratio of $\rm H_2O(2_{02}-1_{11})$/CO$_{8-7}$ seem comparable with those of the nearby 
starburst/enshrouded-AGN Mrk~231, suggesting that SDP.17b could also host a luminous AGN. 
The detection of a strong $\rm H_2O$ $2_{02}-1_{11}$  line in SDP.17b implies an 
efficient excitation mechanism of the water levels that must occur in very dense and
warm interstellar gas. 
}



\maketitle

\section{Introduction}

Gravitational lensed sources have played an important role in infrared and submillimeter studies of high-$z$ galaxies 
since the discovery of IRAS FSC10214+4724 (hereafter IRAS F10214; Rowan-Robinson et al.\ 1991).  The studies of 
this and two other bright strongly-lensed QSOs, APM 08279+5255 (Downes et al.\ 1999) and the Cloverleaf 
(H1413+117; Barvainis et al.\ 1994), demonstrate the utility of using high gravitational magnification 
to investigate the detailed properties of distant galaxies.  These sources allowed pioneering detections of the 
infrared and submillimeter continuum and lines of CO, HCO$^+$, HCN, HNC, CN, CS \& H$_2$O   
(see e.g., Solomon \& Vanden Bout 2005,  Riechers et al.\ 2011a)  
These three lensed sources also provided the first few spatially  resolved measurements on scales of hundreds of parsecs.
Before {\it Herschel} these sources were without peer, except for SMMJ2135--0102, MM~18423+5938 and SXDF1100.001
(Swinbank et al.\ 2010; Lestrade et al.\ 2010; Ikarashi et al.\ 2010).

Now the wide area surveys from space -- especially H-ATLAS and HerMES with
{\it Herschel} which will observe 570\,deg$^2$ and 70\,deg$^2$, respectively
(Eales et al.\ 2010, Oliver et al.\ 2010) -- and from the ground,
for example, with the South Pole Telescope (Vieira et al.\ 2010) are
increasing the area of submillimeter surveys by factors of hundreds over previous surveys.
We expect that the number of strongly lensed submillimeter sources will also increase by a very large factor.
From the analysis and intensive follow up of the five strongest lenses found in the H-ATLAS Science Demonstration
(SDP) field of 14.5 deg$^2$ (Negrello et al.\ 2010, hereafter Ne10), the H-ATLAS team has shown that such lenses
are relatively easy to identify.  With S$_{500\mu m}$ $>$ 100\,mJy and S$_{1.2mm}$ $>\sim$ 10\,mJy, 
their number exceeds that of the unlensed high-$z$  submillimeter sources (Negrello et al.\ 2007).   
The number of similar high-z strongly lensed galaxies in the full H-ATLAS survey will be larger than 100. 

CO lines are almost the only molecular lines currently detected with the
very broad band spectrometers such as Z-Spec (Lupu et al.\ 2010, hereafter Lu10; 
Scott et al.\ 2011) and Zpectrometer (Frayer et al.\ 2011) and
with interferometers (Cox et al.\ 2011, Riechers et al.\ 2011b, Leeuw et al.\ in prep.). 
Besides $\sim$20  CO-line detections in these sources, the only other molecule 
detection currently reported in high-$z$ {\it Herschel} galaxies is the tentative detection 
with Z-Spec of the emission line of para-H$_2$O $2_{02}-1_{11}$ 
(rest LST frequency 987.93\,GHz) in the $z=2.305$ source H-ATLAS J090302.9-014127B -- hereafter 
SDP.17b (Lu10). However, the noise is high in this part of the Z-Spec spectrum and,
in addition, the line is partially blended with another  strong feature 
that was identified as CO(5--4) emission from the intervening massive lensing galaxy at $z=0.942$ 
H-ATLAS J090302.9-014127A -- hereafter SDP.17a -- as discussed in Lu10 and
Ne10. Here we report new measurements of the  H$_2$O 2$_{02}-1_{11}$ line in SDP.17b 
using the IRAM Plateau de Bure interferometer (PdBI) confirming the line and enabling 
a more detailed study of its properties. 

We adopt a cosmology with $H_{0}=71\,{\rm km\,s^{-1}\,Mpc^{-1}}$,
$\Omega_{M}=0.27$, $\Omega_{\Lambda}=0.73$ (Spergel et al.\ 2003).

\begin{figure*}
\centering
\rotatebox{0}{\includegraphics[height=7cm]{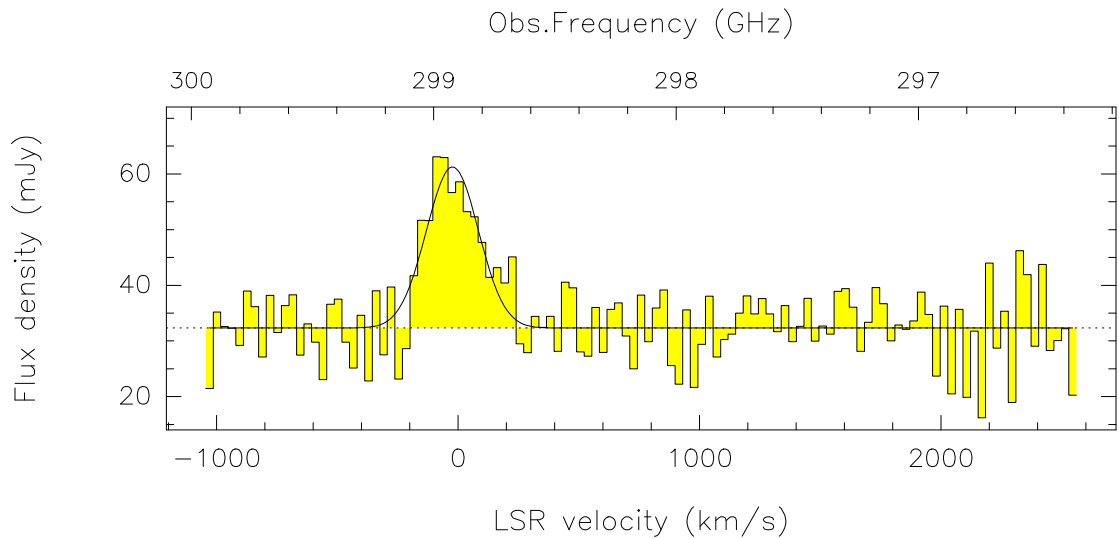}}
\hfill
\caption{Spectrum of the para H$_2$O $2_{02}-1_{11}$ emission line towards SDP.17b, where the velocity 
scale is centered on its observed frequency at 298.93\,GHz (corresponding to $z=2.3049$). The r.m.s. 
noise is $\sim$4.7\,mJy/beam in 31.6\,MHz channels. A Gaussian fit to the H$_2$O spectrum is shown 
as a full line while the dotted line shows the underlying dust continuum emission. The H$_2$O line 
is clearly asymmetric and not well fit by a Gaussian profile.}
\end{figure*}

   \begin{table*}
      \caption[]{Observed parameters of the H$_2$O $2_{02}-1_{11}$ emission line in SPD.17b}
         \label{tab:lines}
            \begin{tabular}{l c c c c c c c c}
            \hline
            \noalign{\smallskip}
           Source & $\nu_{\rm rest}$ & $\nu_{\rm obs}$ & $z$ & S$_\nu$ & $\Delta$V$_{\rm FWHM}$ 
                  & $I$           & L$^a$                & L'$^a$/10$^{10}$  \\
                  & [GHz]           & [GHz]           &          & [mJy]   & [km/s] 
                  & [Jy km s$^{-1}$]&  [10$^7$ L$_{\sun}$] & [K km/s\,pc$^2$] \\
            \noalign{\smallskip}
            \hline
            \noalign{\smallskip}
SDP.17b & 987.93 & 298.93 & 2.3049\,$\pm$\,0.0006 & 29 & 250\,$\pm$\,60 & 7.8\,$\pm$\,0.5 & 85\,$\pm$\,0.6 &  2.5\,$\pm$\,0.2  \\
            \noalign{\smallskip}
            \hline
           \end{tabular}
\begin{list}{}{}
\item[] Quoted errors are the statistical errors derived from a Gaussian fit to the line profile.  
\item[] $^a$Eqs. 1 \& 3 of Solomon et al.\ (1997). The line luminosities are not corrected for the
magnification. Typical amplifications of $\sim$10 (or more) are derived for the
{\it Herschel} detected lensed sources (Ne10).
\end{list}
\end{table*}

\section{Water Lines in High-$z$ Galaxies}

Conducting studies of H$_2$O in high-$z$ galaxies is important. If not locked in grains 
H$_2$O may be one of the most abundant molecules in the gas. It is known to be a tracer of 
the dense, warm gas and possibly of strong infrared radiation because its large dipole and high energy 
levels make its excitation difficult and sensitive to the interstellar conditions (Gonz\'alez-Alfonso 
et al.\ 2010, hereafter G-A10). 

Since {\rm ISO}, far-infrared  H$_2$O lines are known to be prominent in local ULIRGs
and composite AGN/starburst galaxies, such as Mrk~231 and Arp~220, which show  
a series of H$_2$O lines in absorption (Gonz\'alez-Alfonso et al.\ 2008, 2004).  
The recent {\it Herschel} SPIRE FTS submillimeter spectrum of Mrk~231 provides a wealth of
molecular emission lines including high-J CO lines up to J=13, seven rotational lines of $\rm H_2O$ almost 
comparable in strength to the CO emission lines, as well as many other species
(OH$^+$, H$_2$O$^+$, CH$^+$ and HF) which were never observed before in external galaxies 
(van der Werf et al.\ 2010 [hereafter vdW10]; G-A10); Fischer et al.\ 2010). 
A high intensity of the H$_2$O lines implies a high H$_2$O
column density in a compact nuclear component, and thus a H$_2$O abundance
approaching 10$^{-6}$ (G-A10). This is most probably the consequence of
shocks, cosmic rays, dense hot cores, possibly an XDR (X-ray dominated) moderated chemistry,
and/or an 'undepleted chemistry' where grain mantles have evaporated.
Therefore a composite model might be necessary to explain the water
emission -- shocks excited by the mechanical energy of the starburst
or perhaps an AGN may enhance the gas phase abundance of water, while a
strong infrared radiation field from a intense starburst or AGN (similar to
Mrk~231) may be responsible for the high excitation.  
As stressed by G-A10, the high H$_2$O/CO ratio makes it unlikely that the H$_2$O emission originates
in classical Photo-Dissociation Regions (PDR), since in the Orion Bar, the proto-typical 
Galactic PDR, CO lines are a factor $\geq 50$ stronger than the H$_2$O lines. 
Similarly, in M\,82, this ratio is $\approx 40$ (Wei\ss\ et al.\  2010).
The favored model for explaining these extraordinary features implies a high H$_2$O 
abundance and the presence of a small star forming disk, 
composed of clumps of dense gas exposed to strong ultraviolet radiation, dominating 
the emission of CO lines up to $\rm J=8$  (vdW10, G-A10). X-rays from the accreting supermassive 
black hole in Mrk~231 are likely to contribute significantly to the 
excitation and chemistry of the inner disk, as shown by the presence of OH$^+$ and 
H$_2$O$^+$ lines. 

The number density of  galaxies similar to Mrk ~231 is expected to be at least two orders 
of magnitude larger at high than at low redshift (Fabian et al.\ 2000; Alexander et al.\ 2005). They should
represent a significant fraction of all submillimeter galaxies (SMGs), and thus of {\it Herschel} lensed
SMGs. At high-$z$, the far-infrared lines of molecular species such as
$\rm H_2O$ are redshifted into the atmospheric sub/millimeter windows. 
For strongly lensed sources, these molecular lines are within the detection reach of present sub/millimeter 
interferometers such as Plateau de Bure Interferometer (PdBI) and future facilities such as ALMA and NOEMA. 
Tentative detections of H$_2$O were reported in the Cloverleaf for the $2_{20}-1_{11}$ transition 
(Bradford et al.\ 2009, Table 2) and IRAS~F10214 for the $2_{11}-2_{02}$ transition (Casoli et al.\ 1994). 
In addition, a luminous water maser ($\rm \nu_{\rm rest} = 22.2 \, GHz$) was detected in a lensed quasar  
at $z=2.64$ (Impellizzeri et al.\ 2008) and tentatively reported in F10214 (McKean et al.\ 2011). 
The H$_2$O 1$_{10}$ -- 1$_{01}$ line was also detected in absorption  at z\,=\,0.685 towards B0218+357 by Combes \& Wiklind (1997).
It is likely 
that in the near future, the whole set of molecular lines (including the water lines) seen in 
Mrk~231 and other local sources (vdW10) will be detectable with ALMA in high-$z$ lensed galaxies, 
provided they are comparable to Mrk~231.

\section{Observations and Results}

In order to confirm the detection towards SDP.17b of the redshifted  H$_2$O $2_{02}-1_{11}$ emission line, 
we used the PdBI with six antennas and the new 'Band 4' receiver, which covers the frequency range 277-371\,GHz. 
Since the wide-band correlator, WideX, provides a contiguous frequency coverage of 3.6 GHz in dual polarization,  
it allowed us to include the frequency of 297\,GHz at the edge of the bandpass where Lu10  reported a  second strong,
but partially blended line, which they identified as the CO(5--4) emission of the lensing galaxy SDP.17a at 
$z=0.942\pm0.004$.

First observations wer made in the compact D-configuration on 2011 January 3 in conditions with good atmospheric 
phase stability (seeing of 0.7$''$) and reasonable transparency (PWV $\le$0.5\,mm). They were completed by observations 
in extended A- and B-configurations in February and March 2011.  With a total of $\sim 6.2$~hour
on-source integration, a strong signal was detected both in the continuum and in the purported H$_2$O line (Fig.\ 1). 
The dust continuum flux density at 1.0~mm is 32.3$\pm$2\,mJy in good agreement with the value derived 
from Z-Spec by Lu10, and used in the SED fits of Ne10 and Lu10. However, the respective contributions 
of SDP.17b and SDP.17a to this value remain uncertain. The maximum flux density and integrated intensity of 
the H$_2$O line are $\rm 29\,mJy$ and $\rm 7.8\pm0.5\, Jy \, km \, s^{-1}$, respectively, with a FWHM of 
$250\pm60\, \rm km \, s^{-1}$. The line central frequency of 298.93\,GHz corresponds to $z= 2.3049\pm0.0006$. 
Compared to the value previously reported by Lu10 of $19\pm7 \, \rm Jy \, km \, s^{-1}$, the intensity  
is lower by a factor 2.4 and marginally consistent within 1.6\,$\sigma$. The relatively low angular resolution 
($\approx 1^{\prime\prime}$) did not allow us to study the spatial properties of the signal. The source does 
not seem to be resolved and it is unlikely that a significant flux is missed, either from the $\rm H_2O$ line or 
the continuum flux, as shown by the consistency of the latter with the Z-Spec value.  

Clearly, there is no other strong line in this spectral range with an intensity approaching that of 
H$_2$O. In particular, this precludes the presence of a CO(5--4) line of SDP.17a stronger than about 
3\,Jy\,km\,s$^{-1}$, i.e.\ 1/10 of the tentative detection reported by Lu10 (29$\pm$9\,Jy km s$^{-1}$), 
in the redshift range $0.922 < z < 0.944$. However, emission of SDP.17a in the CO(5--4) line is not  
ruled out and could be present at a redshift $z>0.944$, outside of the bandpass of the current observations.

\section{Discussion: High-excitation Gas}

\subsection{Properties of SDP.17b}

In order to assess the implication of the detection of H$_2$O emission in SDP.17b, it is important to 
summarize the current information on the properties of SDP.17b and to replace this source in relation 
with other submillimeter lensed sources. SDP.17b is one of the five prominent high-$z$ lensed SMGs found by 
Ne10 in the H-ATLAS SDP field, with an apparent infrared luminosity $L_{\rm IR} = 4\,10^{13} \, \rm L_{\sun}$ (8--1000\,$\mu$m, Lu10). 
The information on the lensing system remains limited since no high-resolution sub/millimeter image is yet available. 
However, Ne10 identified that the deflector SDP.17a could consist of two foreground lensing masses at similar 
high redshifts ($z \sim 0.8-0.9$). The amplification factor of SDP.17b is still unknown and could reach values of 
$\approx 10$ or more (Ne10). The infrared luminosity of SPD.17b is thus comparable to typical ULIRGs, including 
Mkr~231. The mid-infrared photometry of SPD.17b is unknown, but, as it is detected at $100 \, \mu$m (Ne10), 
there might be an AGN contribution.

   \begin{table}
      \caption[]{Properties of the para H$_2$O $2_{20}-1_{11}$ emission line in active galaxies}
         \label{tab:lines}
            \begin{tabular}{l c c c c}
            \hline
            \noalign{\smallskip}
             Source  &  $z$       &  $I({\rm H_2O})^a$  & $L({\rm H_2O})^a$  &  $I_{\rm H_2O} ^a/I_{\rm CO} ^b$  \\
                     &            &   [Jy km s$^{-1}$]      &  [10$^7$ L$_{\sun}$]    &                                    \\
            \noalign{\smallskip}
            \hline
            \noalign{\smallskip}
SDP.17b             &  2.305     &   7.8\,$\pm$\,0.5    &  85/$\mu _L$      &        0.5$\pm$0.2   \\
Mrk 231$^1$         &  0.042     &	718	        &    2.4 	    &        $\sim$0.5 	   \\ 
Cloverleaf$^2$      &  2.565     &   20.3$\pm$6.1       &  21(11/$\mu _L$)  &        0.4$\pm$0.2    \\
            \noalign{\smallskip}
            \hline
           \end{tabular}
 \begin{list}{}{}
 \item[]  $^a$H$_2$O 2$_{02}$-1$_{11}$ line  
 \item[]  $^bI$(H$_2$O 2$_{02}$-1$_{11}$)/$I$(CO[8--7]) from Lu10
 \item[]  [1] G-A10; [2] Brafdford et al. (2009)
  \end{list}
\end{table}

The most remarkable property of SDP.17 is the richness of its 200-300\,GHz Z-Spec spectrum (Lu10). Besides 
the H$_2$O line, it displays three CO lines (J=6--5, 7--6(+[C{\small I}]) and 8--7) at $z=2.3$. Despite the 
relatively low individual S/N ratios ($\sim 1.5-3$) of these line intensities, their distribution has no clear sign of 
a turnover up to J\,=\,8--7 pointing to the similarity with both Mrk~231 and the Cloverleaf (Lu10).  
In fact, the CO spectral energy distribution of Mrk~231 also peaks in between the CO(5--4) and CO(6--5) transitions, 
but it has in addition a strong high excitation component as seen in vdW10. However, there is yet no information about 
CO lines above J\,=\,8--7 in SDP.17b. Note that SDP.17b is one of the rare examples among the {\it Herschel} strongly lensed 
galaxies (together with  HERMESJ~105751.1+573027 [HLSW-01], Scott et al.\ 2011) that display strong
high-J CO lines. While the information about high-J CO lines is still lacking in many of them, several other 
well studied high-$z$ lensed sources either from H-ATLAS (J091304.9-005344 (SDP.130), Lu10; J142413.9+022304 
(SDP.15.141), Cox et al.\ 2011), or from elsewhere (Wei\ss\ et al.\ 2005; Danielson et al.\ 2011; 
Lestrade et al.\ 2010) have a clear turnover at lower J values.

Nothing is yet published about the results of observations of lower-J CO millimeter lines of SPD.17b. However, several works are in progress at CARMA, GBT/Zpectrometer and IRAM/PdBI, including observations of lines J\,=\,1--0, 3--2 and 4--3 (Leeuw et al.\ in prep.; Frayer et al.\ in prep.; Cox et al.\ in prep.). Their results will provide various information about the molecular gas and the lensing system. In particular, it will be interesting to compare the H$_2$O and CO line profiles and eventually their spatial images, and to study the connection between the warm gas emitting H$_2$O lines and the colder gas emitting CO lines, and especially the cold gas traced by the CO J\,=\,1--0 line (Frayer et al.\ in prep.).

The FIRST radio survey (Becker et al. 1995) yields S$_{\rm 1.4GHz}$ = 464$\pm$145\,$\mu$Jy for SDP.17. From the 
$L_{\rm IR}$ value and using the definition of Sajina et al.\ (2008), we find for the IR/radio parameter $q =2.53$. 
This value is well within the range of $q$ values  for $z \sim 2$ starburst sources (e.g.\ Sajina et al.\ 2008; 
Fiolet et al.\ 2009; Ivison et al.\ 2010). This suggests that SDP.17b is not a radio-loud source and that 
star formation  is responsible for most of its far-infrared luminosity (similar conclusions could apply to SDP.17a).
However, it would be important to check whether the actual radio spectral index could suggest that some fraction 
of the radio emission is powered by an AGN, as in the case of SPD.81  (H-ATLAS J090311.6+003907; Valtchanov et al.\ 2011).

\subsection{Implication of H$_2$O Emission in SDP.17b}

As discussed in \S 2, the detection of H$_2$O in SDP.17b proves special excitation conditions in a warm dense gas and 
an intense infrared radiation field, similar to Mkr~231. Indeed there is currently no information in SDP.17b about 
other strong emission in H$_2$O lines with higher excitation, as in the case of Mrk~231, and it remains difficult 
to infer the detailed conditions of the gas from the detection of a single line. However, despite the uncertainty 
in the amplification factor, it is clear that the intrinsic H$_2$O line luminosity is at least 
comparable to that of Mrk~231 (Table~2). Similarly, the ratio I($\rm H_2O (2_{02}-1_{11})$)/I($\rm CO(8-7)$) 
seems comparable for both sources and completely different from standard PDRs (\S 2). All the evidence 
points to the fact that SDP.17b and Mkr~231 have similar properties. SDP.17b may thus display excitation 
conditions as special as and a chemistry as rich as Mkr~231, suggesting the influence of a luminous AGN.

However, strictly speaking, in the absence of  observation of other transitions, the excitation of H$_2$O in 
SDP.17b could be lower than in Mrk~231 and mostly limited to levels with energy $\sim$\,100\,K, 
as the $2_{02}$ level. As discussed by G-A10, 
such an excitation might be achieved in less extreme conditions
found in warm dense gas of a more extended region, through dense hot
cores, and/or shocks, and is not necessarily associated with excitation by a
powerful AGN.  But it remains unclear if the above conditions
could boost the H$_2$O/CO line intensity ratio to the values that are observed.

We conclude that SDP.17b is likely to be an analogue of Mkr~231. Most of the water lines detected by 
wdW10 in Mkr~231 could already be detected in SDP.17b or in other {\it Herschel} lensed sources 
(including SDP.81 and HLSW-01) using the PdBI today or during the Early Science of ALMA. 

The  results reported in this paper are an example of the studies that can be initiated when many 
bright, lensed high-$z$ submillimeter galaxies become available (through {\it Herschel} or other facilities).
Follow-up observations of these sources, especially with the increased sensitivities afforded by ALMA or NOEMA, 
will allow to undertake comprehensive studies of molecular lines in sources similar to SDP.17b and provide 
new insights into the physical conditions of the dense warm molecular gas of high-$z$ SMGs and their AGN.

\begin{acknowledgements}
Based on observations carried out with the IRAM Plateau de Bure
interferometer. IRAM is supported by INSU/CNRS (France), MPG (Germany)
and IGN (Spain).  The authors are grateful to the IRAM staff for
their support. US participants in H-ATLAS acknowledge support from NASA 
through a contract from JPL. Italian participants in H-ATLAS acknowledge a 
financial contribution from the agreement ASI-INAF I/009/10/0. 
The SPIRE development has been supported by national funding agencies: CSA (Canada); NAOC (China); CEA, CNES,
CNRS (France); ASI (Italy); MCINN (Spain); SNSB (Sweden); STFC (UK); and NASA (USA).

\end{acknowledgements}

\end{document}